\documentstyle[12pt,a4wide]{article}
\input epsf
\newcommand{\be}{\begin{equation}}
\newcommand{\ee}{\end{equation}}
\newcommand{\pr}{\partial}
\newcommand{\I}{{\cal I}}
\font\mybb=msbm10 at 11pt

\def\bb#1{\hbox{\mybb#1}}

\def\bR {\bb{R}}

\newcommand{\news}{\setcounter{equation}{0}}
\def\ben{\begin{equation}}
\def\een{\end{equation}}
\def\bea{\begin{eqnarray}}
\def\eea{\end{eqnarray}}
\begin{document}

\title{\vskip -0pt
\bf \large \bf ICOSAHEDRAL SKYRMIONS\\[30pt]
\author{Richard A. Battye$^{1}$,
Conor J. Houghton$^{2}$ and
Paul M. Sutcliffe$^{3}$\\[10pt]
\\{\normalsize $^{1}$
{\sl Jodrell Bank Observatory, Macclesfield, Cheshire SK11 9DL U.K.}}
\\{\normalsize {\sl $\&$  Department of Physics and Astronomy,
Schuster Laboratory,}}
\\{\normalsize {\sl University of Manchester, Brunswick St,
 Manchester M13 9PL, U.K.}}
\\{\normalsize {\sl Email : rbattye@jb.man.ac.uk}}\\
\\{\normalsize $^{2}$ {\sl School of Mathematics,}}
\\{\normalsize {\sl Trinity College, Dublin 2, Ireland }}
\\{\normalsize {\sl Email : houghton@maths.tcd.ie}}\\
\\{\normalsize $^{3}$  {\sl Institute of Mathematics,
 University of Kent at Canterbury,}}\\
{\normalsize {\sl Canterbury, CT2 7NF, U.K.}}\\
{\normalsize{\sl Email : P.M.Sutcliffe@ukc.ac.uk}}\\}}
\date{October 2002}
\maketitle
 
\begin{abstract}
In this paper we aim to determine the baryon numbers at which the
minimal energy Skyrmion has icosahedral symmetry.
By comparing polyhedra which arise as minimal energy Skyrmions with
the dual of polyhedra that minimize the energy of Coulomb charges on a sphere,
we are led to conjecture a sequence of magic baryon numbers, 
$B=7,17,37,67,97,...$
at which the minimal energy Skyrmion has icosahedral symmetry and 
unusually low energy. We present evidence for this conjecture by applying
a simulated annealing algorithm to compute energy minimizing rational maps
for all degrees upto 40. Further evidence is provided by the explicit
construction of icosahedrally symmetric rational maps of degrees 
37, 47, 67 and 97. 
To calculate these maps we introduce two new methods for computing
rational maps with Platonic symmetries. 

\end{abstract}

\newpage
 
\section{Introduction}\news
Skyrmions  are topological solitons in three space dimensions which
are candidates for an effective description of nuclei, with an identification
between soliton and baryon numbers \cite{Sk}. Recently, the minimal energy
Skyrmions for all baryon numbers $1\le B\le 22$ were computed 
and their symmetries identified \cite{BS}. The baryon density of these
Skyrmions is localized around the vertices and edges of polyhedra, which
are almost always trivalent, and for $B\ge 7$ are composed of 12 pentagons
and $2B-14$ hexagons, with only a few exceptions (which
can be understood by a symmetry enhancement principle). These Skyrmions
have discrete point group symmetries, including occasional Platonic
symmetries. For $B=7$ and $B=17$ the minimal energy Skyrmion is
particularly symmetric, having icosahedral symmetry $Y_h$, and the value
of the energy is unusually low. However, there are other baryon numbers
at which icosahedral Skyrme fields exist, but the minimal energy Skyrmion
has less symmetry.
This motivates the main question addressed
in this paper, namely, what are the magic baryon numbers at which the
minimal energy Skyrmion has icosahedral symmetry with a resulting unusually
 low energy?

To gain some insight into this problem we note that, as first observed in
\cite{AS}, there is a close relationship between the polyhedra which arise
as minimal energy Skyrmions and the duals of polyhedra which occur in the
problem of minimizing point Coulomb charges on a sphere. This latter problem
is often known as the Thomson problem, even though he appears not to have
posed it explicitly, and we shall use this nomenclature here. The Thomson
problem is well studied for upto 200 points on the sphere \cite{MDH} and
generically the $n$ points sit at the vertices of a combinatoric deltahedron.
Taking the dual of a deltahedron leads to a trivalent polyhedron, which is
the class of polyhedra which generically arise for Skyrmions. A Skyrmion
polyhedron with baryon number $B$ has  $2B-2$ faces, so to identify
this with the dual of a Thomson polyhedron requires that we consider
$n=2B-2$ Coulomb charges on the sphere. Let us denote by $G_B$ the
symmetry of the minimal energy Skyrmion with baryon number $B$ 
and by $H_B$ the symmetry
of the minimal energy Thomson configuration of $2B-2$ points on the sphere.
Extracting the information from references \cite{BS} and \cite{Ed} we 
obtain Table 1, in which we compare the symmetries of the minimal energy
 configurations for each problem. 

From Table 1 it is clear that although a variety of different 
Platonic, dihedral and cyclic symmetry
groups occur, there is a remarkable match for the two problems in 17 out of
 the 22
cases. Moreover, a closer inspection reveals that in these 17 cases
not only do the symmetry groups match, but the combinatorial types of
the Skyrmion polyhedron and the dual of the Thomson polyhedron are
identical. The five examples that do not coincide, $B=5,9,10,19,22$
shows that the topography of the two energy functions is slightly different
 and suggests
that the same factors which determine the polyhedron (or its dual) 
are important, but perhaps with slightly different
weightings. For example, for $B=9$ and $B=10$, which are
not particularly low in energy, it is known that Skyrmion configurations
exist which have the symmetries required to match to those of the Thomson
problem, but that they have very slightly higher energy than the minimal energy
Skyrmion. The fact that there is so often an agreement for the two problems
leads us to believe that, in the cases where particularly symmetric low
energy configurations arise, they will be the minima in both problems.
Thus, as a working hypothesis to test,  we shall postulate that $G_B$ is the
 icosahedral group only if $H_B$ is also the icosahedral group.  
\begin{table}
\setlength{\tabcolsep}{2.0pt}
\begin{tabular}{|c|cccccccccccccccccccccc|}
\hline
$B$&1&2&3&4&5&6&7&8&9&10&11&12&13&14&15&16&17&18&19&20&21&22\\ \hline
$G_B$&$O(3)$&$D_{\infty h}$ & $T_d$ &$O_h$ &$D_{2d}$&$D_{4d}$&$Y_h$ & $D_{6d}$
& $D_{4d}$& $D_3$ & $D_{3h}$ & $T_d$ & $O$& $C_2$& $T$ & $D_2$& $Y_h$& $D_2$
& $D_3$& $D_{6d}$ & $T_d$ & $D_3$\\ \hline
$H_B$&$O(3)$&$D_{\infty h}$ & $T_d$ &$O_h$ &$D_{4d}$&$D_{4d}$&$Y_h$ & $D_{6d}$
& $T$& $D_{4d}$ & $D_{3h}$ & $T_d$ & $O$& $C_2$& $T$ & $D_2$& $Y_h$& $D_2$
& $D_2$& $D_{6d}$ & $T_d$ & $D_{5h}$\\ \hline
\end{tabular}
\caption{The symmetry group $G_B$ of the minimal energy Skyrmion
with baryon number $B$ and the symmetry group $H_B$ of
the $2B-2$ points which minimize the Thomson problem.}
\label{tab-sym}
\end{table}
\normalsize
Although the Thomson problem is a difficult one to study numerically,
it is certainly much easier than finding minimal energy Skyrmions, so
we can take advantage of the known numerical results. Minimal energy
configurations are currently available for upto 200 points, 
that is, $B\le 101$,
and of these the values $B=7,17,37,62,67,97$ are selected as magic
numbers at which the configuration has icosahedral symmetry and unusually
low energy when compared to a numerical fit of all 200 configurations 
\cite{MDH}.
As we describe later, the case $B=62$ is rather different from the 
others in the sequence, so we shall leave this example out for the moment.
We are thus led to conjecture that there is a sequence
$B=7,17,37,67,97,...$
at which the minimal energy Skyrmion has icosahedral symmetry and 
unusually low energy. In the rest of this paper we perform some investigations
to test this conjecture, and hence the connection between Skyrmions and
the Thomson problem.

\section{Minimizing Rational Maps}\news
A static Skyrme field, $U({\bf x}),$ is an $SU(2)$ matrix defined
throughout $\bR^3$ and satisfying the boundary condition that 
$U \rightarrow 1$ as $|{\bf x}|\rightarrow \infty.$
This boundary condition
implies
a compactification of space so that the Skyrme field becomes a mapping 
$U:S^3\mapsto SU(2),$ and so can be classified by an integer valued winding
number
\be
B=\frac{1}{24\pi^2}\int \epsilon_{ijk}\,\mbox{Tr}\left(\pr_i U\,
U^{-1}\pr_j U\, U^{-1}
\pr_k U\, U^{-1}\right)d^3x,\label{bar}
\ee
which is a topological invariant. This winding number counts the number of
solitons in a given field configuration and is identified with  baryon number
in the application to modelling nuclei.

The energy of a static Skyrme field is given by
\be
E=\frac{1}{12\pi^2}\int\mbox{Tr}\left(-\frac{1}{2}\left(\pr_iU\,
U^{-1}\right)^2-\frac{1}{16}\left[\pr_iU\,
U^{-1},\;\pr_j U\, U^{-1}\right]^2\right)d^3x
\label{energy}
\ee
and for each integer $B,$ the problem is to minimize this energy within
the class of fields with baryon number $B$ in order to find the minimal
energy Skyrmion. This problem has been solved numerically for all $B\le 22$
\cite{BS} yielding the results presented in Table 1 for the symmetries
of the minimal energy Skyrmions. When we refer to the symmetry of a Skyrmion
we do not mean that the Skyrme field itself is invariant under particular
spatial rotations, but rather that the effect of a spatial rotation can
be undone by the application of the global $SO(3)$ symmetry of the Skyrme
model, which acts through the conjugation
  $U\mapsto {\cal O}U{\cal O}^\dagger,$
where ${\cal O}\in SU(2)$ is a constant matrix. In particular this means that
the baryon and energy densities (the integrands in (\ref{bar}) and 
(\ref{energy})) are strictly invariant.
 
It is computationally prohibitive
to apply the full numerical scheme to larger values of $B$ at the present time,
but fortunately an approximation method has been developed which provides
very accurate results. This is the rational map ansatz \cite{HMS}, 
where a Skyrme field with baryon number $B$ is constructed from a degree
$B$ rational map between Riemann spheres. Although this
ansatz does not give exact solutions of the static Skyrme equations,
it produces approximations which have energies only a few percent
above the numerically computed solutions.
Briefly, use spherical coordinates in $\bR^3$, so
that a point ${\bf x}\in\bR^3$ is given  by a pair $(r,z)$, where
$r=\vert{\bf x}\vert$ is the  distance from the origin, and $z$ is a
Riemann sphere coordinate giving the point on the unit two-sphere
which intersects the half-line through the origin and the point ${\bf
x}$.
Now, let $R(z)$ be a degree $B$ rational map between Riemann spheres,
that is, $R=p/q$ where $p$ and $q$ are polynomials in $z$ such that
$\max[\mbox{deg}(p),\mbox{deg}(q)]=B$,  and $p$ and $q$ have no common
factors.  Given such a rational map the ansatz for the Skyrme field is
\be  U(r,z)=\exp\bigg[\frac{if(r)}{1+\vert R\vert^2} \pmatrix{1-\vert
R\vert^2& 2\bar R\cr 2R & \vert R\vert^2-1\cr}\bigg]\,,
\label{rma}
\ee where $f(r)$ is a real profile function satisfying the  boundary
conditions $f(0)=\pi$ and $f(\infty)=0$, which is determined by
minimization of the Skyrme energy of the field (\ref{rma}) given a
particular rational map $R$. 

Substitution of the rational map ansatz (\ref{rma}) into the Skyrme
energy functional results in the following expression for the energy \be
E=\frac{1}{3\pi}\int \bigg( r^2f'^2+2B(f'^2+1)\sin^2 f+\I\frac{\sin^4
f}{r^2}\bigg) \ dr\,,
\label{rmaenergy}
\ee where $\I$ denotes the integral \be \I=\frac{1}{4\pi}\int \bigg(
\frac{1+\vert z\vert^2}{1+\vert R\vert^2}
\bigg\vert\frac{dR}{dz}\bigg\vert\bigg)^4 \frac{2i \  dz  d\bar z
}{(1+\vert z\vert^2)^2}\,.
\label{i}
\ee To minimize the energy (\ref{rmaenergy}) one first
determines the rational map which minimizes $\I$, 
then given the minimum value of $\I$ it is a simple exercise to find the
minimizing profile function.
Thus, within the rational map ansatz, the problem of finding
the minimal energy Skyrmion reduces to the simpler problem of
calculating the rational map which minimizes the function
$\I$.
 
The baryon density of the rational map $R=p/q$ is proportional to the Wronskian
\be
w(p,q)=p'q-q'p
\label{wron}
\ee
which has $2B-2$ roots, giving the points on the Riemann sphere for
which the baryon density vanishes along the corresponding half-lines
through the origin. These $2B-2$ points on the sphere give the face-centres
of the Skyrmion polyhedron, or equivalently the vertices of the dual 
polyhedron which is associated with the Thomson problem.

Using a simulated annealing algorithm the $\I$ minimizing rational maps
for $1\le B\le 22$ have been computed \cite{BS} and found to be in
good agreement with the results of full Skyrme field minimization.
Here we extend the simulated annealing computation to $B\le 40,$ in an
attempt to determine particularly low energy magic numbers. In the Thomson
problem the magic numbers are determined by comparing the energy of
minimal solutions with the energy of a numerical fit to the data of
 all known minimal energy solutions - thereby isolating cases where the
energy is lower than the expected fit. In the problem of minimizing rational
maps it turns out that there is a more natural approach, due to the fact that
a useful lower bound exists. Using a simple inequality it is shown in \cite{HMS}
that $\I\ge B^2.$ It turns out that examining the excess above this bound,
by computing the quantity $\I/B^2$, is a good diagnostic tool for highlighting
low energy maps, and in particular is more useful than simply calculating the
energy of the associated Skyrme field. We illustrate this in Fig.~1 by plotting
$\I/B^2$ for the results of our simulated annealing computations for
$2\le B\le 40.$
\begin{figure}[ht]
\begin{center}
\leavevmode
\vskip -3cm
\epsfxsize=15cm\epsffile{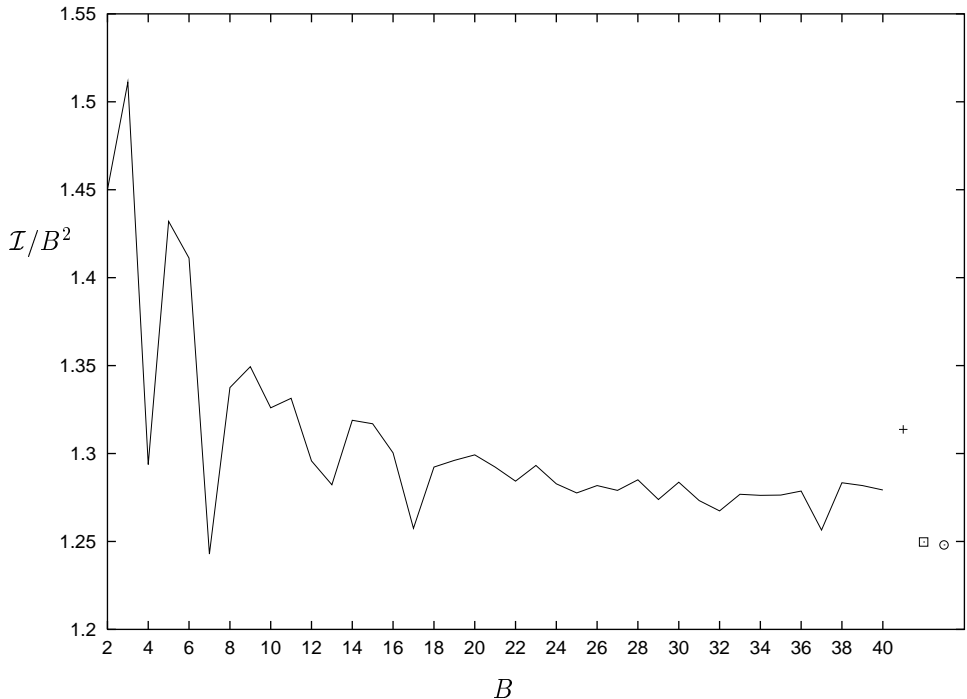}
\vskip -8.5cm
\caption{$\I/B^2$ for $2\le B\le 40,$ as calculated 
by $\I$ minimizing simulated annealing computations.
Also shown are the values of $\I/B^2$ for the icosahedrally
symmetric Skyrmions with $B= 47$ (cross), $B=67$ (square)
and $B=97$ (circle), discussed in the text.}
\label{fig1}
\end{center}
\end{figure}
There are clear dips at the magic numbers
 $B=7$ and $B=17,$ corresponding to the already known low energy icosahedral 
Skyrmions (see Table 1). There are also dips at $B=4$ and $B=13$, where
it is known that the Skyrmions also have Platonic symmetry, but this time
octahedral (see Table 1). However, there is one more major dip at $B=37,$ and this
 is precisely the value predicted as the next magic number in the icosahedral
 sequence suggested by comparison
 with the Thomson problem. This result, therefore,
provides strong support for our conjectured sequence, providing we can prove
that this low energy degree 37 map obtained from simulated annealing does
indeed have icosahedral symmetry. This will be discussed in the next section.

Note that the quantity $\I/B^2$ appears to be
tending towards an asymptotic value of around $1.28,$ apart from magic numbers
where it drops to around $1.25.$ It would be interesting to understand this
approach to a relatively constant value, as well as its magnitude.
\begin{figure}[ht]
\begin{center}
\leavevmode
\vskip -4cm
\epsfxsize=18cm\epsffile{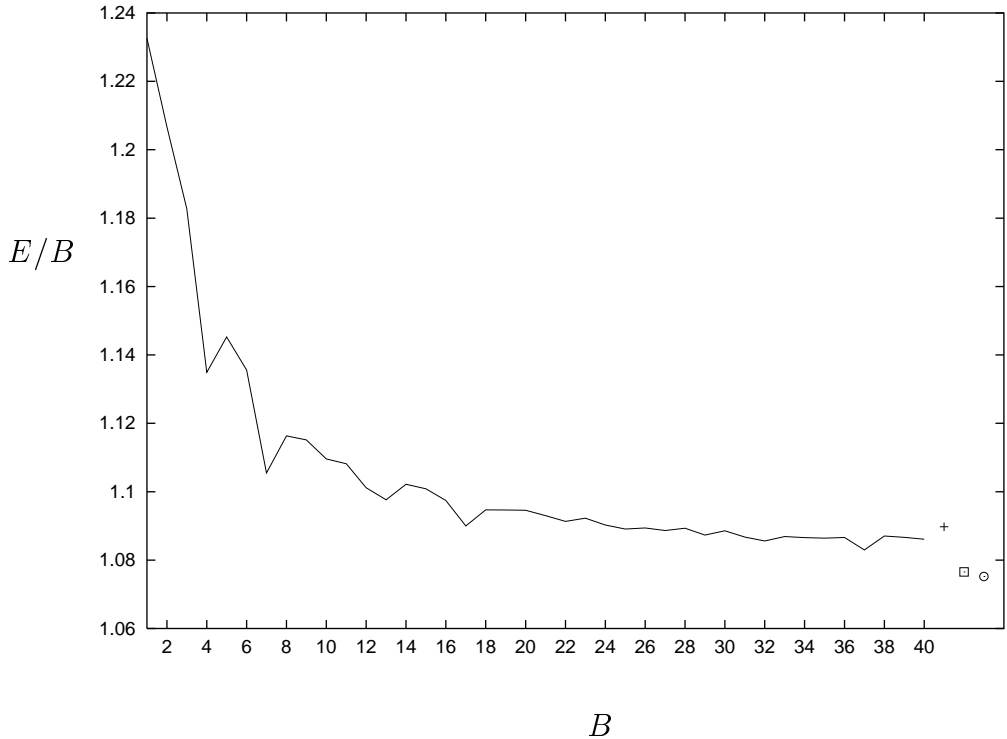}
\vskip -11.5cm
\caption{The energy per baryon $E/B$ of
the Skyrmions constructed from the minimizing maps with $1\le B\le 40.$ 
Also shown are the values of $E/B$ for the icosahedrally
symmetric Skyrmions with $B= 47$ (cross), $B=67$ (square)
and $B=97$ (circle), discussed in the text.}
\label{fig2}
\end{center}
\end{figure}

In Fig.~2 we plot the energy per baryon $E/B$ of
the Skyrmions constructed from the minimizing maps with $1\le B\le 40.$ 
The dips at the magic numbers are clearly visible in Fig.~2, reproducing
the sequence displayed in Fig.~1. However, in this case the dips are superimposed upon
a general decrease of $E/B$ with increasing $B,$ which is why we regard
the quantity $\I/B^2$ as a more useful diagnostic than $E/B.$ 

The radius of a Skyrmion produced from the rational map ansatz can be defined
as the radial value, $r_*,$ at which the profile function is equal to $\pi/2.$
In Fig.~3 we plot $r_*^2$ as a function of $B$ for $1\le B\le 40.$
This clearly shows that the radius has a  $\sqrt{B}$ dependence, and given that
all these configurations are reasonably close to the Faddeev-Bogomolny energy bound
$E\ge |B|$, this means that the energy grows like the square of the radius, as 
expected for a shell-like structure. Note that at the magic numbers the 
radius of the Skyrmion is slightly less than expected, presumably due to
a more compact arrangement of a particularly symmetric energy density.
\begin{figure}[ht]
\begin{center}
\leavevmode
\vskip -4cm
\epsfxsize=18cm\epsffile{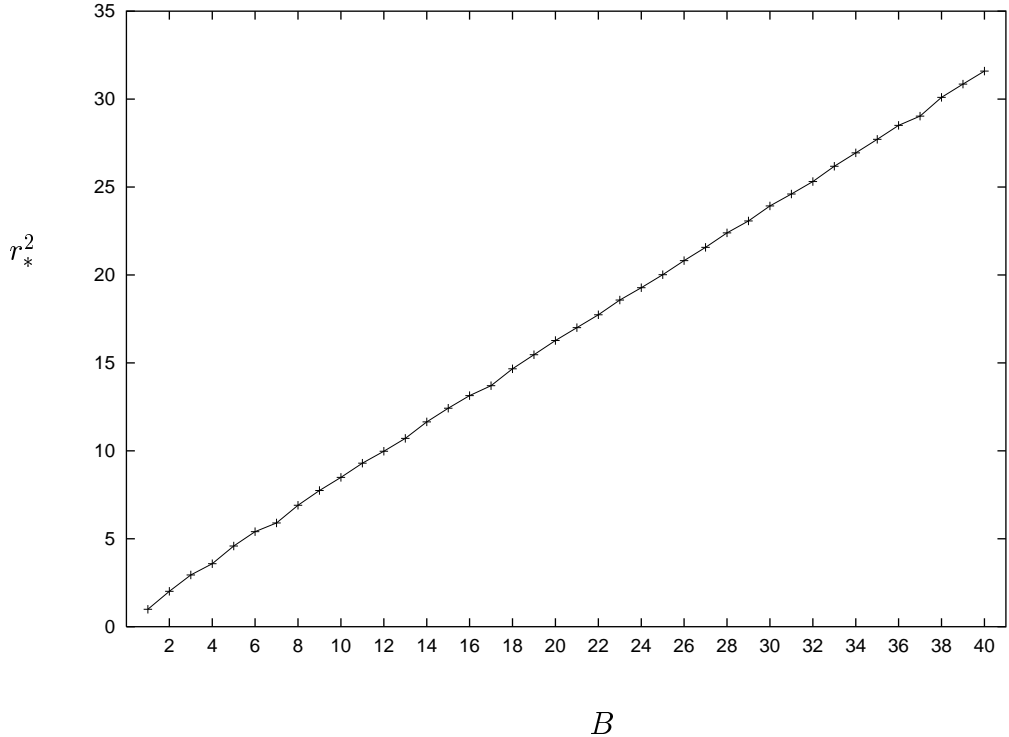}
\vskip -11.5cm
\caption{$r_*^2,$ the squared radius of the Skyrmion, as a function
of baryon number $B.$}
\label{fig3}
\end{center}
\end{figure}
\section{Computing Icosahedral Maps}\news
Recall that a Skyrmion is symmetric under a group $G\subset SO(3),$
if a spatial rotation $g\in SO(3)$ can be compensated by an action of
the global $SO(3)$ symmetry. In terms of the rational map approach 
a spatial rotation acts on the Riemann
sphere coordinate $z$ as an $SU(2)$ M\"obius transformation.
Similarly the global symmetry acts as on the Riemann
sphere coordinate $R$ of the target 2-sphere  as an
$SU(2)$ M\"obius transformation.
Hence, a map is $G$-symmetric if, for each  $g\in G$, there exists a target
space rotation $D_g$ such that $R(g(z))=D_g(R(z)).$
Since we are dealing with $SU(2)$ transformations the set of target
space rotations will form a representation of the double group of $G$,
but we shall continue to call this $G.$

To determine the existence and compute particular symmetric rational
maps is a matter of classical group theory. We are
concerned with degree $B$ polynomials which form the carrier space for
$\underline{B+1}$, the $(B+1)$-dimensional  irreducible representation
of $SU(2).$ Now, as a representation of $SU(2)$ this is irreducible,
but if we only consider the restriction to a subgroup $G$,
$\underline{B+1}\vert_G$, this will in general be reducible. What we
are interested in is the irreducible decomposition of this
representation and tables of these subductions  can be found, for
example, in ref.~\cite{pgtt}.

The simplest case in which a $G$-symmetric degree $B$ rational map
exists is if  \be \underline{B+1}\vert_G=E+...  \ee where $E$ denotes
a two-dimensional representation. Here, and in the following, the dots
denote representations of dimension higher than those shown. 
In this case a basis for $E$
consists of two degree $B$ polynomials which can be taken to be the
numerator and denominator of the rational map.  A subtle point which
needs to be addressed is that the two basis polynomials may have a
common root, in which case the resulting rational map is degenerate
and does not correspond to a genuine degree $B$ map.

More complicated situations can arise, for example, if \be
\underline{B+1}\vert_G=A_1+A_2+...  \ee  where $A_1$ and $A_2$ denote
two one-dimensional representations, then a whole one-parameter family
of maps can be obtained by taking a constant multiple of the ratio of
the two polynomials which are a basis for $A_1$ and $A_2$
respectively.  An $m$-parameter family of $G$-symmetric maps can be
constructed if the decomposition contains $(m+1)$ copies of a
two-dimensional representation, that is, \be
\underline{B+1}\vert_G=(m+1)E+... \ee where the $m$ (complex)
parameters correspond to the freedom in the choice of one copy of $E$
from $(m+1)E.$

A detailed explanation of how to explicitly calculate any required symmetric
map is given in \cite{HMS}. However, this approach involves computing
appropriate projectors which are matrices of size $(B+1)\times(B+1),$ and
even with the use of symbolic computational packages this procedure becomes
cumbersome for the large values of $B$ that we are interested in here.
In this section we therefore describe and apply two new, more convenient,
 methods for calculating symmetric maps. We shall concentrate on the
situation of relevance to this paper, where $G=Y,$ the icosahedral group,
 but the methods are applicable for any $G.$

Before we describe our new approaches we need to recall some facts about 
representations of the icosahedral group and Klein polynomials of the
icosahedron. 
The icosahedral group $Y$ has the trivial 1-dimensional representation
$A$ and two 2-dimensional representations, which we denote by $E_1'$ and $E_2',$
with a prime denoting the fact that these are representations of the
double group of $Y$ which are not representations of $Y.$ There are also
three, four, five and six-dimensional representations, but we shall not need
these here.

Klein polynomials are strictly invariant polynomials for the Platonic groups
\cite{Kl}.
Since,
\be \underline{13}|_Y=A+...\ee
this implies that there is degree 12 invariant polynomial. This is the
Klein polynomial given by
\be k_v=z^{11}+11z^6-z\ee
and although it appears to have degree 11, it should be thought of as
having degree 12 with one root at infinity. The
roots of this polynomial, considered as points on the Riemann sphere,
are located at the vertices of a suitably oriented and scaled icosahedron.
The same construction, but using the face-centres and mid-points of the
edges of the icosahedron, in place of the vertices, produces the Klein
polynomials
\bea k_f&=&z^{20}-228z^{15}+494z^{10}+228z^5+1 \\
k_e&=&z^{30}+522z^{25}-10005z^{20}-10005z^{10}-522z^5+1\eea
which are also $Y$-invariant, by construction. 

\subsection{Polarization}
In this subsection we describe our polarization method for computing
symmetric maps. It has similar features to the polarization technique
used to construct symmetric Nahm data \cite{HMM}.

Suppose we wish to obtain the symmetric degree $B$ map associated with the
decomposition
\be \underline{B+1}|_Y=E+...\label{dc1}\ee
where $E$ denotes one of the 2-dimensional representations.
The above fact implies that 
\be E\otimes\underline{B+1}|_Y=A+...\label{dc2}\ee
and, in our polarization method, the invariant polynomial corresponding to
this 1-dimensional representation is used to construct a basis for the $E$
in (\ref{dc1}).

It is convenient to work with homogeneous coordinates $x,y$ on the Riemann
sphere, that is, $z=x/y$, so that a polynomial in $z$ of degree $B$ corresponds
to a homogeneous degree $B$ polynomial in $x$ and $y.$

Let $(p_L(x,y),q_L(x,y))$ be known degree $L$ polynomials which form a basis
for the representation $E$ and let $k(x,y)$ be a degree
$B+L$ invariant polynomial which is a basis for the 1-dimensional representation
in (\ref{dc2}). Then, since the pair $(\partial_y,-\partial_x)$ transform
in the same way as the pair $(x,y)$ under linear $SU(2)$ transformations,
this means that the polynomials $p_B(x,y),q_B(x,y)$ defined by
\be
p_B(x,y)=p_L(\partial_y,-\partial_x)k(x,y), \quad
q_B(x,y)=q_L(\partial_y,-\partial_x)k(x,y)\label{polar}
\ee
have degree $B$ and are the required basis for the 2-dimensional
representation occuring in (\ref{dc1}). 

As an example of this scheme we now construct the icosahedrally 
symmetric degree 17 rational map, in an orientation that we shall require
later. The relevant decomposition is
\be \underline{18}|_Y=E_2'+...\ee
so we first require a known $Y$-symmetric rational map that is a basis
for the representation $E_2'.$ The simplest known example is the
degree 7 map \cite{HMS} (corresponding to the $B=7$ $Y_h$-symmetric
minimal energy Skyrmion)
\be
p_7(x,y)=x^7-7x^5y^2-7x^2y^5-y^7, \quad
q_7(x,y)=x^7+7x^5y^2-7x^2y^5+y^7.
\ee
Hence we have $B=17$ and $L=7,$ so now we require an invariant
polynomial $k(x,y)$ of degree $B+L=24.$ This is easily found by using
an appropriate combination of Klein polynomials, in this case
$k(x,y)=k_v^2(x,y)$ is the degree 24 invariant
where $k_v(x,y)=x^{11}y+11x^6y^6-xy^{11}$ is the degree 12 Klein polynomial
given earlier, when written in terms of homogeneous coordinates.
The formula (\ref{polar}) then produces
\bea
p_{17}&=&z^{17}+17z^{15}+119z^{12}-187z^{10}+187z^7+119z^5+17z^2-1\nonumber\\
q_{17}&=&z^{17}-17z^{15}+119z^{12}+187z^{10}+187z^7-119z^5+17z^2+1\label{Y17}
\eea
when written in terms of the inhomogeneous coordinate $z.$
This map is equivalent, after a change of spatial and internal orientation,
to the $Y_h$-symmetric map presented in \cite{HMS} that corresponds to
the  $B=17$ $Y_h$-symmetric minimal energy Skyrmion. In the following
subsection we shall require this map in the orientation presented in 
(\ref{Y17}).

Although this method is much easier to implement than the projector algorithm,
it turns out that for the icosahedral maps we require in this paper there
is yet another approach, which is even more effecient.

\subsection{Klein Leapfrog}

From the previous section we already have two $Y$-symmetric rational
maps, which are $R_7=p_7/q_7$ and $R_{17}=p_{17}/q_{17}.$
Here we describe how the other $Y$-symmetric maps that we require can be
obtained from these two by the simple multiplication of
 invariant Klein polynomials. This way of obtaining  higher
degree invariant rational maps we refer to as the Klein Leapfrog method. 

Recall that we wish to determine whether the low energy $\I$ minimizing map
of degree 37 that we found earlier is icosahedrally symmetric.
The relevant decomposition is
\be \underline{38}|_Y=E_1'+2E_2'+...\label{dc3}\ee
Both the maps $(p_7,q_7)$ and $(p_{17},q_{17})$ are a basis for the
representation $E_2'$, and the multiplication of these maps by any
(integer power of a) Klein polynomial does not change the transformation
properties, since Klein polynomials are invariants.
Thus, both $(k_ep_7,k_eq_7)$ and $(k_fp_{17},k_fq_{17})$ are degree
37 $Y$-symmetric maps. Each map alone is not a valid degree 37 rational map,
since the numerator and denominator contain common factors, but taken together
they form an acceptable basis for the $2E_2'$ in (\ref{dc3}). 
Explicitly,
\be R_{37}=\frac{p_{37}}{q_{37}}=\frac{k_fp_{17}+ck_ep_7}{k_fq_{17}+ck_eq_7}
\label{Y37}\ee
where $c$ is a complex parameter. For $c=0$ or $c=\infty$ the
map is clearly degenerate, having degree lower than 37, but for 
generic values of $c$ the numerator and denominator are coprime.

The Wronskian of this map must be strictly invariant, and indeed it
is given by the following combination of Klein polynomials
\be
w(p_{37},q_{37})=k_e^2k_v(80c+28c^2)-k_f^3k_v(68+120c).
\ee
The 72 roots of this polynomial give the face-centres of the Skyrmion polyhedron.
Minimizing the integral $\I$ over the one (complex) parameter family of maps
(\ref{Y37}) results in a minimum at $c=-0.829+0.545i,$ where $\I/B^2=1.255.$
This is precisely the value found by the minimization over all degree 37 maps,
so we confirm that the $\I$ minimizing degree 37 map, and hence the minimal
energy $B=37$ Skyrmion, has icosahedral symmetry. Note that the symmetry
group is only $Y$ and not $Y_h$ since $c$ is not real. A baryon density
isosurface plot for the associated Skyrmion is displayed in Fig.~4a.
\begin{figure}[ht]
\begin{center}
\leavevmode
\vskip -0cm
\epsfxsize=15cm\epsffile{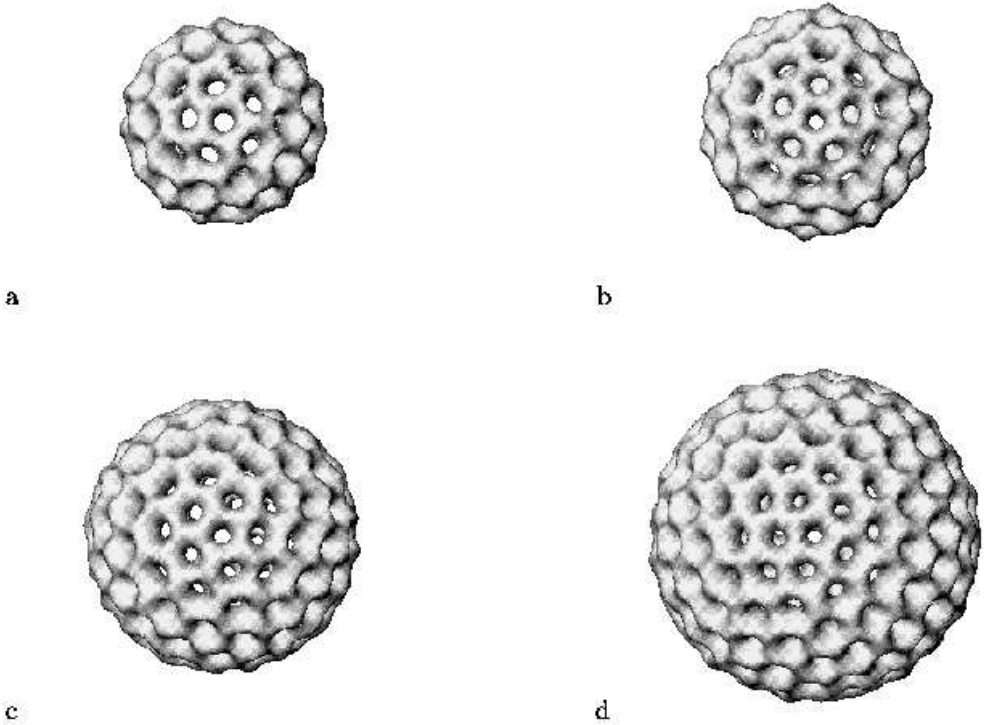}
\vskip -0cm
\caption{Baryon density isosurface plots (to scale) for icosahedral
Skyrmions with baryon numbers a) $B=37$, b) $B=47$, c) $B=67$, d) $B=97$.}
\label{fig4}
\end{center}
\end{figure}

The next magic number in our conjectured icosahedral list is $B=67.$
The application of our simulated annealing scheme to extend the 
results presented in Fig.~1 to such a large value of $B$ would require
unreasonable computing resources. We therefore make use of the fact that
the quantity $\I/B^2$ appears to approach an asymptotic value of around
$1.28,$ whereas for icosahedral magic numbers the value is closer to $1.25.$
Therefore we aim to present evidence in support of our conjecture by
finding an icosahedral map of degree 67 with $\I/B^2\approx 1.25.$

The relevant decomposition is
\be
\underline{68}|_Y=2E_1'+3E_2'+...
\ee
so we require three degree 67 maps to form a basis for the second
component in the above decomposition. These are given by a Klein leapfrog as
\be
(k_ek_fp_{17},k_ek_fq_{17}), \quad (k_e^2p_7,k_e^2q_7), \quad
(k_f^3p_7,k_f^3q_7).
\ee
It might seem strange that $(k_v^5p_7,k_v^5q_7)$ is not included, but
because $\underline{61}|_Y=2A+...,$ there must be a linear relationship
between $k_e^2$, $k_f^3$ and $k_v^5.$ In fact, it is given by
$1728k_v^5=k_e^2-k_f^3,$ \cite{Kl}.

Minimizing over the two (complex) parameter family of maps
\be
R_{67}=\frac{k_ek_fp_{17}+c_1k_e^2p_7+c_2k_f^3p_7}
{k_ek_fq_{17}+c_1k_e^2q_7+c_2k_f^3q_7}
\ee
yields a minimum at $c_1=-0.292-0.816i,$ $c_2=-0.491+1.008i,$ for
which $\I/B^2=1.250.$ This value is plotted as the square in Fig.~1,
and it is clearly consistent with being an icosahedral magic number,
as is the energy per baryon of the associated Skyrmion which is 
plotted as the square in Fig.~2.
A baryon density isosurface derived from the minimal $Y$-symmetric
map is displayed in Fig.~4c.

The next magic number on our list is $B=97.$ The required decomposition
is
\be
\underline{98}|_Y=3E_1'+4E_2'+...
\ee
so there is a three (complex) parameter family of $Y$-symmetric maps.
Three of the required basis maps are obtained by a Klein leapfrog of the
three degree 67 basis maps given above, through the multiplication by $k_e.$
The fourth basis map is a Klein leapfrog of $(p_{17},q_{17})$ through
the multiplication by $k_f^4.$ The full map is therefore
\be
R_{97}=\frac{k_e^2k_fp_{17}+c_1k_e^3p_7+c_2k_ek_f^3p_7+c_3k_f^4p_{17}}
{k_e^2k_fq_{17}+c_1k_e^3q_7+c_2k_ek_f^3q_7+c_3k_f^4q_{17}}.
\ee
A minimization over the three complex parameters yields a minimum
for $c_1=0.705+0.699i, c_2=-0.554-1.701i, c_3=-0.967+0.930i,$
 at which $\I/B^2=1.248.$ 
This value is plotted as the circle in Fig.~1, and again it is consistent
with being the minimal degree 97 map, producing another icosahedral
magic number at $B=97.$ 
A baryon density isosurface of the Skyrmion derived from this minimal
 $Y$-symmetric map is displayed in Fig.~4d. The energy per baryon
of this Skyrmion is plotted as the circle in Fig.~2. Given that the rational
map ansatz tends to overestimate the energy by around one or two percent, then
the true energy per baryon of this Skyrmion must be very close to that of the hexagonal
lattice \cite{BS2}, which has $E/B=1.061.$

Icosahedrally symmetric rational maps, and hence Skyrme fields, exists
for many values of $B,$ but rarely are these symmetric configurations
those of minimal energy. The simplest example is the degree 11 rational
map presented in \cite{HMS}. For this map $\I/B^2=3.84,$ which is clearly
very large, and indeed the associated $B=11$ Skyrme field has larger
energy than 11 well-separated single Skyrmions. This is not very surprising,
given that the associated polyhedron is an icosahedron - clearly violating
the favourable trivalent property at all vertices. However, in the Thomson
problem there are more subtle examples, where there is an icosahedrally
symmetric configuration which has reasonably low energy, but not quite
as low as a less symmetric minimal energy solution. This situation occurs
for the values $B=22,47,82,...$ \cite{MDH}. We shall see if this situation
is also mirrored in the Skyrmion problem, by studying $Y$-symmetric rational
maps of degree $47.$

The required decomposition is
\be
\underline{48}|_Y=E_1'+2E_2'+...
\ee
and a basis for the $2E_2'$ is obtained by the Klein leapfrog of $R_{17}$
by $k_e$ and the Klein leapfrog of $R_7$ by $k_f^2.$ Therefore, the
one parameter family of $Y$-symmetric maps is
\be
R_{47}=\frac{k_ep_{17}+ck_f^2p_7}{k_eq_{17}+ck_f^2q_7}.
\ee
Minimizing over $c$ yields a minimum when $c$ is real
(so the symmetry extends to $Y_h$) and takes
the value $c=-1.425,$ at which $\I/B^2=1.314.$ This value is plotted as
the cross in Fig.~1, and it can be seen that, even though it is reasonably
low, it is not consistent with the general trend for minimal energy
maps. The associated energy per baryon is plotted as the cross in Fig.~2
and provides further evidence that this is  not a minimal energy Skyrmion.
This suggests that the same phenomenon of non-minimal icosahedral 
maps exists in both the Thomson and Skyrme problems, providing yet more
evidence for the similarity of these two systems.
A baryon density isosurface is displayed in Fig.~4b for the $Y_h$-symmetric
Skyrmion obtained from the above map with the minimal value of $c.$
From this figure it can be seen that the Skyrmion polyhedron fits into
the required class, as a trivalent polyhedron with 12 pentagonal faces
and the remaining faces hexagonal. Therefore, the reason for it not to be
the minimal energy Skyrmion must be subtle, and probably involves the
placement of the pentagons within the polyhedron, when compared to a more
favourable but less symmetric distribution.

Finally, we turn to the anomalous case of $B=62.$ 
In the Thomson problem there is an icosahedral magic number at $B=62$,
but the relevant decomposition for rational maps is
\be
\underline{63}|_Y=A+\,\mbox{irreps of dimension greater than 2}
\ee
so there is certainly no $Y$-symmetric degree 62 rational map, and
probably no $Y$-symmetric $B=62$ Skyrmion either. The resolution of this
problem is the fact that the polyhedron associated with a Skyrmion is
derived from the baryon density, and it is possible that the baryon
and energy density of a Skyrmion could have more symmetry than the 
Skyrme field itself. In terms of the rational map ansatz this corresponds
to an enhanced symmetry of the Wronskian, not shared by the rational
map.   

Recall that the Wronskian is a polynomial of degree $2B-2,$ so to
see if this is a possible explanation for the case $B=62$ we
need to look for $Y$-invariant polynomials of degree 122.
The decomposition
\be
\underline{123}|_Y=2A+...
\ee
reveals that there are two invariants, and in fact they are given by
$k_vk_fk_e^3$ and $k_vk_f^4k_e.$ Thus, to address this case we would need to
find the family of degree 62 rational maps, $(p_{62},q_{62})$
 so that the Wronskian takes the form
\be
w(p_{62},q_{62})=k_vk_fk_e(c_1k_e^2+c_2k_f^3)
\ee
where $c_1$ and $c_2$ are arbitrary complex constants. It is difficult to
see how to explicitly construct this family, given we do not know the
 symmetry of the rational map, but it would be interesting if this could be
done, to see whether a map with a $Y$-invariant Wronskian is likely to
be the minimal map. However, as far as our definition of icosahedral
magic number is concerned, the value $B=62$ does not qualify because
the map is not $Y$-symmetric.

The example of a non-minimal icosahedral $B=22$ Thomson configuration,
briefly mentioned above, also appears to fall into the same class~\cite{BS2}.
There are no $Y$-symmetric degree 22 maps, but there is a degree 42
invariant, given by $k_vk_e,$ to which the Wronskian of a degree 22
map could be proportional.

\section{Conclusion}\news
In this paper we have used a comparison between Skyrmion polyhedra
and the duals of Thomson polyhedra to predict a sequence of
magic baryon numbers at which the Skyrmion has icosahedral symmetry
and unusually low energy. We have presented some evidence for our
conjecture, through the minimization of the most general rational
maps for all degrees upto 40, and by the explicit construction,
using two new methods, of some high degree rational maps with
icosahedral symmetry.

Our methods could also be used to find other possible minimal energy
rational maps and Skyrme fields, with octahedral and tetrahedral symmetries.
It is likely that these other Platonic symmetries are more prevalent than
icosahedral symmetry, and may account for some of the less pronounced dips
in Fig.~1.

Finally, a comparison between the Skyrme crystal and the Skyrme lattice
\cite{BS2} suggests that for large enough baryon numbers the shell-like
structure of Skyrmions may give way to a crystal structure. However, even
the order of magnitude of $B$ at which this transition might take place
is not known, so whether all the icosahedral Skyrmions we have constructed
will survive this possible transition remains an open question.

\section*{Acknowledgements}
We acknowledge EPSRC (PMS) and PPARC (RAB) for Advanced Fellowships.


\begin{thebibliography}{100}

\bibitem{pgtt} S.L. Altmann and P. Herzig, \lq{\sl
Point-Group Theory Tables}\rq, Oxford, Clarendon Press, 1994.

\bibitem{AS} M.F. Atiyah and P.M. Sutcliffe, 
Proc. Roy. Soc. A 458, 1089 (2002).

\bibitem{BS} R.A. Battye and P.M.Sutcliffe, Phys. Rev. Lett. 79, 363 (1997);
 Phys. Rev. Lett. 86, 3989 (2001); Rev. Math.
 Phys. 14, 29 (2002).

\bibitem{BS2} R.A. Battye and P.M. Sutcliffe, Phys. Lett. B 416, 385 (1998).

\bibitem{Ed}
J.R. Edmundson, Acta Cryst. A48, 60 (1992).

\bibitem{HMM} N.J. Hitchin, N.S. Manton and M.K. Murray,
 Nonlinearity 8, 661 (1995).

\bibitem{HMS} C.J. Houghton, N.S. Manton and P.M. Sutcliffe,
Nucl. Phys. B {510}, 507 (1998).

\bibitem{Kl} F. Klein, \lq{\sl Lectures on the icosahedron}\rq,
London, Kegan Paul, 1913.

\bibitem{MDH} J.R. Morris, D.M. Deaven and K.M. Ho, 
Phys. Rev. B 53, R1740 (1996).

\bibitem{Sk} T.H.R. Skyrme, Proc. Roy. Soc. A 260, 127 (1961).

\end{thebibliography}
\end{document}